\documentclass[11pt,twoside]{article}
\usepackage{amssymb}
\usepackage{latexsym}
\usepackage{slashed,cite}
\usepackage{lscape}
\evensidemargin\oddsidemargin
\newcommand{\Red}[1]{#1}
\newcommand{\OliveGreen}[1]{#1}
\newcommand{\RawSienna}[1]{#1}
\newcommand{\Blue}[1]{#1}
\newcommand{\Maroon}[1]{#1}
\newcommand{\ft}[2]{{\textstyle\frac{#1}{#2}}}
\def\Re{\mathop{\rm Re}\nolimits}
\def\Im{\mathop{\rm Im}\nolimits}

\def\rmi{{\rm i}}
\def\rmd{{\rm d}}
\newcommand{\hc}{{\rm h.c.}}
\newsavebox{\uuunit}
\sbox{\uuunit}
    {\setlength{\unitlength}{0.825em}
     \begin{picture}(0.6,0.7)
        \thinlines
        \put(0,0){\line(1,0){0.5}}
        \put(0.15,0){\line(0,1){0.7}}
        \put(0.35,0){\line(0,1){0.8}}
       \multiput(0.3,0.8)(-0.04,-0.02){12}{\rule{0.5pt}{0.5pt}}
     \end {picture}}
\newcommand {\unity}{\mathord{\!\usebox{\uuunit}}}
\newcommand{\bbox}{\lower.2ex\hbox{$\Box$}}

\csname @addtoreset\endcsname{equation}{section}
\newcommand{\SU}{\mathop{\rm SU}}
\newcommand{\SO}{\mathop{\rm SO}}
\newcommand{\U}{\mathop{\rm {}U}}
\newcommand{\USp}{\mathop{\rm {}USp}}

\newcommand{\Symp}{\mathop{\rm {}Sp}}
\newcommand{\Sl}{\mathop{\rm {}S}\ell }

\newcommand{\bitem}[1]{\vspace{-2mm} 
\begin{#1} \setlength{\itemsep}{-5pt} } 
\newcommand{\eitem}[1]{\end{#1}\vspace{-2mm}} 

\begin{document}

\label{xyzt}
\leavevmode\vadjust{\vskip -50pt}

{\noindent\sl
Proceedings of the XI Fall Workshop on Geometry and Physics,\\
Oviedo, 2002\\
\rm
Publicaciones de la RSME, vol.~xxx,
}
\null\vskip 10mm

%
%

\pagestyle{myheadings} \markboth {{\small\sc Supergravity theories}}
{{\small\sc Antoine Van Proeyen}}

%
%

\thispagestyle{empty}
\begin{flushright}
KUL-TF-03/01\\
hep-th/0301005
\end{flushright}

\begin{center}
{\Large\bf Structure of supergravity theories\footnote{Work supported in
part by the European Community's Human Potential Programme under contract
HPRN-CT-2000-00131 Quantum Spacetime} } \vskip 10mm

%
%

{\large\bf Antoine Van Proeyen } \vskip 5mm {\it
Instituut voor Theoretische Fysica, Katholieke Universiteit Leuven,\\
       Celestijnenlaan 200D B-3001 Leuven, Belgium.

\smallskip

email:  Antoine.VanProeyen@fys.kuleuven.ac.be }
\end{center}

\bigskip

%
%

\begin{abstract}
\parindent0pt\noindent
We give an elementary introduction to the structure of supergravity
theories. This leads to a table with an overview of supergravity and
supersymmetry theories in dimensions 4 to 11. The basic steps in
constructing supergravity theories are considered: determination of the
underlying algebra, the multiplets, the actions, and solutions. Finally,
an overview is given of the geometries that result from the scalars of
supergravity theories.



\end{abstract}

%
%
\tableofcontents
\section{Introduction}

Supergravity is now mostly known as an ingredient of superstring theory,
the theory that tells that the elementary particles are vibrations of a
fundamental string. This theory offers a magnificent framework to study
all particles and all interactions. Since 1994 there is an extended
framework connecting all the superstring theories and supergravities,
which is called M-theory. Many unexpected connections now emerge every
year from the study of superstrings and supergravities, increasing
continuously our capabilities to construct realistic models, to perform
calculations on gauge theories, to discuss models of cosmology and to
connect many different aspects of our knowledge of physics and
mathematics.

For many applications of string theory we just need to know the structure
of supergravity theories. The elementary particles that emerge from any
superstring theory and that are not hidden to us by being extremely
massive, nicely fit in supergravity theories. In this review, we will
consider supergravity in its own right, i.e.\ independent of its role in
superstring theory. We will consider the supergravities with fields that
occur in the action up to two derivatives. This corresponds only to the
first order in an $\alpha '$ perturbation theory of superstrings.

The part of the theory that describes the spinless fields is very
illustrative for the structure of such a supergravity theory. These
scalar fields determine the vacua of the theory. In the context of
superstring theory they are the moduli of deformations of the
compactifying dimensions. The scalar fields define a geometry, the
structure of which determines also a large part of the full action due to
the supersymmetry relations. We will at the end of this review consider
these geometries. It turns out that the richest geometric structures are
those that appear when there are 8 supersymmetries.

In the next section, we will review the basic ingredients of supergravity
theory at the bosonic side: the fields that are mostly $n$-forms with
gauge transformations, and dualities between their field strengths. The
mechanism of compensating fields will be shown and we will show how
Poincar{\'e} gravity is obtained as a gauge theory.

The fermionic side is introduced in section~\ref{ss:Clifford}. The
possibilities for supersymmetries are determined by a table of Clifford
algebras with specific properties. The classification of these lead to a
map of possible supergravities and supersymmetric theories that is
explained in section~\ref{ss:mapsusy}.

We can divide the discussion on supergravity theories in 4 steps. First,
in section~\ref{ss:salgebra}, we consider the algebra of the
transformations that leave the action invariant. We discuss the
super-Poincar{\'e} algebra, but also generalizations as central charges, the
(anti-)de Sitter algebra, the conformal algebra and their supersymmetric
extensions. Apart from the spacetime symmetries and the supersymmetries,
other bosonic gauge symmetries play a role. In particular there is the
R-symmetry, for which the supersymmetries form a non-trivial
representation. At the end of this section we show how these algebras are
used in constructing supergravity theories

The second step (section~\ref{ss:multiplets}) brings in the fields as
representations of the algebra. Such a set of fields is a
\emph{multiplet}, and there is a balance between bosonic and fermionic
fields with an \emph{on-shell counting}, but often also with an
\emph{off-shell counting}. We first explain this difference, and then
discuss the most important multiplet in various dimensions and with
various supersymmetry extensions.

In section~\ref{ss:actions} we look at the third step: the construction
of an action and its properties. We explain the duality transformations
in 4 dimensions, and how the kinetic terms of the scalars determine a
scalar geometry. We summarize how isometries generalize to hidden
symmetries of the action, often related to duality transformations, and
how a subgroup of them can be gauged.

The final step (section~\ref{ss:solutions}) involves the properties of
some solutions of the supergravity theories. The concept of BPS solutions
will appear, with charges that are at the boundary of an allowed domain.

As mentioned, the scalars of supergravity theories define a geometry, and
their kinetic terms in the action determine a metric. Often these are
symmetric spaces. An interesting subclass of supergravities, those with 8
supersymmetries, define \emph{special geometries}, which can be real,
K{\"a}hler or quaternionic manifolds. These geometric aspects are reviewed in
section~\ref{ss:scalGeom}, before concluding with some final remarks in
section~\ref{ss:final}.

\section{The ingredients: fields and gauge symmetry} \label{ss:bosIngred}

The bosonic fields that occur in supergravity theories are mostly
antisymmetric tensors (i.e. components of $n$-forms), with gauge
symmetries. There are exceptions, but for most theories one can restrict
to these types of bosonic fields. We will repeat the relevant general
formulae for gauge theories. These fields determine massless particles.
Masses for fields with spin 1 or higher appear by considering
spontaneously broken symmetries. The construction of Poincar{\'e} gravity as
a gauge theory needs already a special treatment.

Symmetries in field theories should be distinguished in different
categories. The simplest ones are \emph{rigid symmetries}. These are
transformations of the fields with a parameter, say $\Lambda $, that is
the same everywhere in spacetime. E.g.\ a rotation of the reference frame
is such a transformation. The second type are the \emph{local symmetries}
or \emph{gauge symmetries}, where this parameter can be taken differently
in any point of spacetime: $\Lambda (x)$, where $x$ denotes the spacetime
point. Furthermore, one should say what is left invariant under
`symmetries'. One can consider symmetries of the action, symmetries of
field equations (transforming a field equation in a linear combination of
field equations), or symmetries of solutions. A symmetry of the action
will also be a symmetry of the field equations but the reverse is not
necessarily true. Such symmetries are not necessarily symmetries of a
solution, and that is the concept of a \emph{spontaneously broken
symmetry}.

We now restrict ourselves to symmetries of the action. When we consider a
rigid symmetry, and try to promote it to a local symmetry, the action $S$
will in general not be invariant due to terms that show the
$x$-dependence of the parameter, i.e.\ $\delta S$ is proportional to
$\partial _\mu \Lambda (x)$,
\begin{equation}
  \delta (\Lambda ) S =\int \rmd^Dx\, J^\mu\,\partial _\mu \Lambda (x)\,,
 \label{delSJdLambda}
\end{equation}
for a $J^\mu $ (`Noether current'). $\mu=0,1,\ldots ,D-1$ labels
spacetime coordinates. To compensate these terms, one needs a field
$A_\mu (x)$ (gauge field) with
\begin{equation}
  S_{\rm new}= S - \int\rmd^Dx\, J^\mu\, A_\mu \,,\qquad
  \delta A_\mu(x) =\partial _\mu \Lambda (x)\,.
 \label{SwithA}
\end{equation}
This is the first step in an iterative procedure to construct invariant
actions (`Noether procedure'). The simplest example of a gauge field is
the Maxwell field that appears when promoting phase transformations of
complex fields that are proportional to their electric charges, to a
local symmetry. The gauge field induces a force between the particles
that transform under the symmetry.

Apart from the scalar fields, all bosonic fields that appear here are
gauge fields. The Maxwell field above can be seen as a 1-form $A=A_\mu
\rmd x^\mu $, and its transformation is $\delta A=\rmd \Lambda$. The
other bosonic fields will be $n$-forms $A^{(n)}$ with a symmetry
parameter that is an $n-1$ form: $\delta A^{(n)}=\rmd \Lambda^{(n-1)}$.
Not all the components of $\Lambda^{(n-1)}=\frac{1}{(n-1)!} \Lambda _{\mu
_1\ldots \mu _{n-1}}\rmd x^{\mu _1}\ldots \rmd x^{\mu _{n-1}}$ are
independent symmetries, as all transformations where
$\Lambda^{(n-1)}=\rmd \Lambda ^{\prime\,(n-2)}$ are not effective.
Counting the remaining gauge-invariant components of an $n$ form gives
\begin{equation}
  {D\choose n} - {D \choose n-1} + {D \choose n-2} - \ldots = {D-1\choose
  n}\,,
 \label{compnform}
\end{equation}
the number of components of an $n$-form in $D-1$ dimensions. This is a
general feature: the independent components form a representation of
$\SO(D-1)$. If furthermore the field equations of massless fields are
used, the independent components form representations of $\SO(D-2)$. The
former is called \emph{off-shell counting}, while the latter is called
\emph{on-shell counting}.  The number of off-shell components is also
equal to the number of components of a massive field. Indeed, the massive
physical fields are also characterized by $\SO(D-1)$ representations, as
that is the little group for a massive state, while $\SO(D-2)$ is the
little group of massless states.

The gauge invariant degrees of freedom for an $n$ form reside in an
$(n+1)$-form field strength $F^{(n+1)}=\rmd A^{(n)}$. This obviously
satisfies a Bianchi identity $\rmd F^{(n+1)}=0$, while, for the standard
Lagrangian proportional to $F^{\mu _1\ldots \mu _{n+1}}F_{\mu _1\ldots
\mu _{n+1}}$,  the field equation gives $\rmd {}^*F=0$. This shows that
${}^* F$ satisfies the same equations as a Bianchi identity for a
$(D-n-1)$-form field strength. Thus, it can be considered as the field
strength of a $(D-n-2)$-form. This shows the \emph{duality} between
$n$-forms and $(D-n-2)$-forms. E.g.\ in 4 dimensions this shows that a
2-form gives physically the same as a scalar, and we can thus restrict
ourselves to scalars and vectors. However, the arguments above are only
true for the simplest actions, i.e.\ with abelian gauge fields. In
non-abelian field theories, antisymmetric tensors can lead to
non-equivalent theories. The duality transformations on vectors is a
\emph{self-duality} between the components of the field strengths $F_{\mu
\nu }$ and transforms electric in magnetic components.

The usual gauge symmetries with 1-form gauge fields are the most
important ingredients of supergravity theories. For a set of gauge
transformations labelled by an index $A$, we write the transformations as
$\delta_A (\Lambda ^A)$ where $\Lambda ^A(x)$ denote the parameters. They
form an algebra as
\begin{equation}
  [\delta_A( \OliveGreen{\Lambda_1^A}),\delta_B(\OliveGreen{\Lambda_2^B})] =\delta_C\left(
\OliveGreen{\Lambda_2^B\Lambda_1^A} \Red{f_{AB}{}^C}\right).
 \label{commutator}
\end{equation}
We discuss here bosonic transformations, but the formulas are also valid
when the parameters are fermionic, i.e.\ for supersymmetries. We need a
gauge field for any symmetry: $\Blue{A_\mu{}^A}$, which appear in
\emph{covariant derivatives}
\begin{equation}
 \RawSienna{D_\mu} =\partial _\mu -\delta_A(\Blue{A_\mu ^A})\,, \qquad
  \delta(\OliveGreen{\Lambda}) \Blue{A_\mu{}^A}=
\partial_\mu\OliveGreen{\Lambda^A} +\OliveGreen{\Lambda^C} \Blue{A_\mu{}^B} \Red{f_{BC}{}^A}\,.
 \label{covder}
\end{equation}
The covariant derivatives involve a sum over all the symmetries.
Replacing ordinary derivatives with these covariant derivatives
eliminates the $\partial \Lambda $ terms in the variation of the action
that were discussed higher. The commutator between these covariant
derivatives give rise to \emph{curvatures}, which are the \emph{field
strength} 2-forms that we mentioned\footnote{(Anti)symmetrization is
always made with `total weight' 1, which means that
$\partial_{[\mu}A_{\nu]}=\ft12\left(
\partial _\mu A_\nu -\partial _\nu A_\mu \right) $. A similar
symmetrization is indicated by $(\mu \nu )$.}:
\begin{equation}
  [\RawSienna{D_\mu},\RawSienna{D_\nu}]=-\delta _A(\RawSienna{R_{\mu \nu
  }{}^A})\,,\qquad
\RawSienna{R_{\mu\nu}{}^A} =
 2\partial_{[\mu}\Blue{A_{\nu]}}{}^A+ \Blue{A_\nu{}^C A_\mu{}^B}
 \Red{f_{BC}{}^A}\,.
 \label{curvatures}
\end{equation}

So far we considered only massless vectors. If we want to describe
massive vectors, we will obtain them from a combination of a massless
vector with a scalar using symmetry breaking. This scalar is then called
a \emph{compensating scalar}. E.g.\ consider a scalar with simple
Lagrangian
\begin{equation}
  {\cal L}=-\ft12 \partial _\mu \Red{\phi}\,\partial ^ \mu\Red{\phi}\,.
 \label{Lscalar}
\end{equation}
There is a symmetry $\delta(\Lambda) \Red{\phi} =M\OliveGreen{\Lambda}$,
where $M$ is a number with dimensions of mass and $\Lambda $ is the
parameter, so far spacetime-independent. If we want to promote this to a
local symmetry, we introduce a gauge field $\Blue{A_\mu}$, and define the
covariant derivative $\RawSienna{D_\mu \phi} =\partial _\mu \Red{\phi}
-M\Blue{A_\mu}$ according to~(\ref{covder}). Replacing the ordinary
derivative by this covariant derivative, the action becomes gauge
invariant, and we can add then also gauge-invariant kinetic terms for the
vector, such that
\begin{equation}
  {\cal L}=-\ft14 \RawSienna{F_{\mu \nu }F^{\mu \nu }}-\ft12\RawSienna{D_\mu \phi D^\mu \phi}
  \,.
 \label{L10}
\end{equation}
As we can now change the scalar to an arbitrary value by the gauge
transformation, we can put it to an arbitrary constant value. This gauge
fixing reduces the action to
\begin{equation}
  {\cal L}=-\ft14 \RawSienna{F_{\mu \nu }F^{\mu \nu }}-\ft12M^2\Blue{A_\mu
  A^\mu}\,,
 \label{Lmassive1}
\end{equation}
the standard action for a massive vector. The procedure of `gauge-fixing'
can also be seen as a redefinition of the gauge field $\Blue{A_\mu }$ to
$\Blue{A_\mu }-\frac{1}{M}\partial _\mu \Red{\phi}$. This is a general
fact: gauge fixing can also be described as a field redefinition. In
principle, the gauge symmetry is never broken, but acts only on the field
that is not present anymore in the action after this redefinition.

If we try to mimic the gauge theory procedure for gravity, we should
consider the Poincar{\'e} group consisting of translations $\OliveGreen{P_a}$
and Lorentz rotations $\OliveGreen{M_{ab}}$. The indices $a,b$ also run
from 0 to $D-1$ as the coordinates of the spacetime manifold, but
indicate vectors or tensors in the tangent space. Each of these should
get a gauge field, leading to the scheme
\begin{equation}
 \begin{array}{|c|c|}\hline
  \OliveGreen{P_a} & \OliveGreen{M_{ab}} \\ \hline
  \Blue{e_\mu{} ^a} & \Blue{\omega _\mu{}^{ab} }\\ \hline
\end{array}
 \label{gravitygaugef}
\end{equation}
$\Blue{e_\mu{} ^a}$ will be the vielbein in general relativity, an
invertible $D\times D$ matrix, and $\Blue{\omega _\mu{}^{ab} }$ the spin
connection. However, in Einstein's theory, the spin connection is not an
independent field. Therefore we impose a covariant constraint: imposing
the vanishing of the translation curvature, defined according
to~(\ref{curvatures}):
\begin{equation}
  \RawSienna{R_{\mu \nu }(P^a)}=2\partial_{[\mu}
\Blue{e_{\nu]}^a}+2\Blue{\omega_{[\mu}{}^{ab}e_{\nu] b}}=0\,.
 \label{constrRP}
\end{equation}
This is called a \emph{conventional constraint}, because it
\textbf{defines} one of the fields, here the spin connection
$\Blue{\omega _\mu{}^{ab} }$. This is now a function of the vielbein with
the expression known in Einstein's theory. 
The transformation of $\Blue{\omega _\mu{}^{ab} }$ is modified with
respect to the one that would directly follow from~(\ref{covder}). You
can see this either from the fact that the transformation of the spin
connection now follows from its new definition, or from the fact that the
transformation should be adapted such that~(\ref{constrRP}) is invariant.
This deforms the algebra. The result is very nice: the translation
operation is replaced by a general coordinate transformation on all
fields. In this way a gauge theory of general coordinate transformations
is obtained and this is Einstein's theory.
\section{The guide: Clifford algebra representations} \label{ss:Clifford}
We now turn our attention to the fermions. Therefore we first of all have
to know which Clifford representations we can use. This information
determines also the amount of supersymmetry that can be present in any
dimension.

What do we have to know? Essentially we need the answers to 3 questions:
 \bitem{enumerate}
  \item How large are the smallest spinors in each dimension?
  \item What are the reality conditions?
  \item Which bispinors are (anti)symmetric?
 \eitem{enumerate}
The latter question is important to know what can occur in a
superalgebra. Supersymmetries are transformations with a spinor parameter
$\epsilon $. E.g.\ a scalar field $\phi (x)$ transforms in a fermion
$\psi (x)$ depending on this parameter:
\begin{equation}
  \delta (\epsilon )\phi(x) =\bar \epsilon\, \psi (x)\,.
 \label{susyscalar}
\end{equation}
The algebra of supersymmetry is that a commutator of two such
transformations leads to general coordinate transformations:
\begin{equation}
  \left[\delta (\epsilon _2)\delta (\epsilon _1)\right] \phi (x)=\bar \epsilon _1\gamma ^\mu
\epsilon _2\partial _\mu \phi(x)= \xi ^\mu\partial _\mu \phi(x)\,.
 \label{susyalgscalar}
\end{equation}
For consistency, the bispinor $\xi^\mu =\bar \epsilon _1\gamma ^\mu
\epsilon _2$ should be antisymmetric in the interchange of $\epsilon_1$
and $\epsilon _2$.

The analysis of gamma matrices and spinors in different dimensions can be
found in many places. We refer especially
to~\cite[section~3]{VanProeyen:1999ni}, which has all the arguments in
the conventions that are used here. Of course, that material is not
original, and is rather a convenient reformulation of earlier works.
Another recent approach to the theory of spinors has been presented
in~\cite{D'Auria:2000ec}, which is convenient also for understanding the
supersymmetry algebras. We now summarize the relevant results.

A priori, a spinor $\Psi $ has $2^{\mathop{\rm Int}\nolimits[D/2]}$
components. One can define a product of all $\gamma $-matrices $\gamma
_*\equiv (-\rmi)^{\mathop{\rm Int}\nolimits[D/2]+1}\gamma _0\gamma
_1\ldots \gamma _{D-1}$, which squares to $\unity $. Note that, though
in~\cite{VanProeyen:1999ni} the formulae occur for an arbitrary number of
timelike dimensions, here we assume always just one timelike direction,
i.e.\ the gamma matrices satisfy
\begin{equation}
  \gamma _{(\mu} \gamma _{\nu)}=
  \ft12\left( \gamma _\mu \gamma _\nu +\gamma _\nu \gamma _\mu \right)
   =\eta _{\mu \nu }=\mathop{\rm diag}\nolimits (-1,+1,\ldots ,+1)\,.
 \label{defgamma}
\end{equation}
For odd dimensions $\gamma _*=\pm \unity $, but for even dimensions it is
nontrivial, allowing non-trivial left and right projections
\begin{equation}
  P_L=\ft12(\unity +\gamma _*)\,,\qquad P_R=\ft12(\unity -\gamma _*)\,.
 \label{PLPR}
\end{equation}
\emph{Weyl spinors} are the reduced spinors under such a projection, thus
e.g.\ a left-handed spinor, which satisfies $P_L\Psi =\Psi $. This
concept (`chirality') thus only makes sense in even dimensions.

One may consider the possibilities for reducing the complex spinors to
spinors that satisfy a reality condition $\Psi ^*=B\Psi $ for a suitable
matrix $B$. This should be consistent with the Lorentz algebra, which is
only possible in some dimensions (and this depends also on the spacetime
signature). The corresponding reduced spinors are called \emph{Majorana
spinors}. If this is not possible, one needs a doublet of spinors to
define reality conditions. In that case the reality conditions do not
really reduce the number of independent components of a spinor, but the
formulation with doublets of real spinors shows more explicitly the
symmetry structure. These spinors are denoted as \emph{symplectic
Majorana spinors}. The reality condition is not always compatible with
the Weyl projection. If the reality condition can be imposed on Weyl
spinors, the corresponding irreducible representations of the Lorentz
algebra are denoted as \emph{Majorana-Weyl spinors} and have only
$2^{D/2-1}$ components.

This leads to table~\ref{tbl:spinors},
\begin{table}[htbp]
  \caption{\it Irreducible spinors, number of components and symmetry
  properties.}\label{tbl:spinors}
\begin{center}
  \begin{tabular}{|c|c|c|c|}
\hline
 Dim & Spinor & min \# components & antisymmetric \\
\hline
 2 & MW & 1 & 1 \\
 3 & M & 2 & 1,2 \\
 4 & M & 4 & 1,2 \\
 5 & S & 8 & 2,3 \\
 6 & SW & 8 & 3 \\
 7 & S & 16 & 0,3 \\
 8 & M & 16 & 0,1 \\
 9 & M & 16 & 0,1 \\
 10 & MW & 16 & 1 \\
 11 & M & 32 & 1,2 \\
\hline
\end{tabular}
\end{center}
\end{table}
where according to the number of dimensions it is indicated whether
Majorana (M), Majorana-Weyl (MW) symplectic (S) or symplectic-Weyl (SW)
spinors can be defined, and the corresponding number of components of a
minimal spinor is given (the table is for Minkowski signature and has a
periodicity of 8). In the final column is indicated which bispinors are
antisymmetric, e.g.\ a 0 indicates that $\bar \epsilon _2\epsilon
_1=-\bar \epsilon _1\epsilon _2$, and a 2 indicates that $\bar \epsilon
_2\gamma _{\mu \nu }\epsilon _1=-\bar \epsilon _1\gamma _{\mu \nu
}\epsilon _2$. This entry is modulo 4, i.e. a 0 indicates also a 4 or 8
if applicable. For the even dimensions, when there are Weyl-like spinors,
the symmetry makes only sense between two spinors of the same chirality,
which occurs for bispinors with an odd number of gamma matrices in these
dimensions $D=2$~mod~4. In the other even dimensions, $D=4$~mod~4, there
are always two possibilities for reality conditions and we give here the
one that includes the `1' as this is the most useful one for
supersymmetry in view of~(\ref{susyalgscalar}).

Consider as an example supersymmetry in 5 dimensions. The fact that `1'
does not appear in the list of antisymmetric bispinors implies that we
cannot have an algebra as in~(\ref{susyalgscalar}). We need anyway for
the reality conditions a doublet of spinor parameters $\epsilon ^i$,
$i=1,2$ at least. The algebra can be of the form
\begin{equation}
  \left[\delta (\epsilon _2),\delta (\epsilon _1)\right] =\bar \epsilon _1^i\gamma ^\mu
\epsilon _2^j\varepsilon _{ij}\partial _\mu\,,
 \label{susyalgD5}
\end{equation}
where now the antisymmetric tensor $\varepsilon _{ij}$ cares for the
antisymmetry between the two parameters. We call this the $N=2$ theory to
indicate the inherent symmetry $\USp(2)=\SU(2)$ between the supercharges,
though it is the simplest one that we can have in 5 dimensions.

 \landscape
\begin{table}[htbp]
  \caption{{\it Supersymmetry and supergravity theories in dimensions 4 to
  11.}  An entry represents the possibility to have supergravity
theories in a specific dimension $D$ with the number of supersymmetries
indicated in the top row. We first repeat for every dimension the type of
spinors that can be used. Every entry allows different possibilities.
Theories with more than 16 supersymmetries can have different gaugings.
Theories with up to 16 (real) supersymmetry generators allow `matter'
multiplets. The possibility of vector multiplets is indicated with
$\heartsuit$. Tensor multiplets in  $D=6$ are indicated by
$\diamondsuit$. Multiplets with only scalars and spin-$\ft12$ fields are
indicated with $\clubsuit$. At the bottom is indicated whether these
theories exist only in supergravity, or also with just rigid
supersymmetry.}
  \label{tbl:mapsusy}
\begin{center}\tabcolsep 5pt
  \begin{tabular}{| *{14}{c|} }
\hline
 $D$ & susy & \multicolumn{4}{c|}{32} & \multicolumn{2}{c|}{24}  & 20 & \multicolumn{2}{c|}{16}  & 12 & 8 & 4  \\
\hline
11  & M & M & \multicolumn{3}{c|}{ } &\multicolumn{2}{c|}{ } &  & \multicolumn{2}{c|}{ }  &  &  &  \\
10  & MW & IIA & IIB & \multicolumn{2}{c|}{ }&\multicolumn{2}{c|}{ } &  &
\begin{tabular}{c}
  I \\
\phantom{$N$}$\heartsuit$ \\
\end{tabular} &  &  &  &  \\
9  & M &  \multicolumn{2}{c|}{$N=2$ } &\multicolumn{2}{c|}{ }&
\multicolumn{2}{c|}{ }  &  &
\begin{tabular}{c}
  $N=1$ \\
\phantom{$N$}$\heartsuit$ \\
\end{tabular} &  &  &  &  \\
8  & M &  \multicolumn{2}{c|}{$N=2$ }&\multicolumn{2}{c|}{ }&
\multicolumn{2}{c|}{ }  &  &
\begin{tabular}{c}
  $N=1$ \\
\phantom{$N$}$\heartsuit$ \\
\end{tabular}&  &  &  &  \\
7  & S &  \multicolumn{2}{c|}{$N=4$ } &\multicolumn{2}{c|}{
}&\multicolumn{2}{c|}{ }  &  &
\begin{tabular}{c}
  $N=2$ \\
\phantom{$N$}$\heartsuit$ \\
\end{tabular}& &  &  &  \\
6  & SW & \multicolumn{2}{c|}{$(2,2)$}&$(3,1)$&$ (4,0)$  &$(2,1)$ &
$(3,0)$& &  \begin{tabular}{c}
  $(1,1)$ \\
\phantom{$N$}$\heartsuit$ \\
\end{tabular} &\begin{tabular}{c} $(2,0)$\\
\phantom{$N$}$\diamondsuit$ \\
\end{tabular}  &  &\begin{tabular}{c} $(1,0)$ \\
$\heartsuit,\diamondsuit,\clubsuit$ \\
\end{tabular}   &  \\
5  & S &  \multicolumn{4}{c|}{$N=8$ }  &\multicolumn{2}{c|}{$N=6$ }  & &
\multicolumn{2}{c|}{
\begin{tabular}{c}
  $N=4$ \\
\phantom{$N$}$\heartsuit$ \\
\end{tabular}}&  &\begin{tabular}{c}  $N=2$ \\
\phantom{$N$}$\heartsuit,\clubsuit$ \\
\end{tabular}  &  \\
4  & M &  \multicolumn{4}{c|}{$N=8$ }  & \multicolumn{2}{c|}{$N=6$ }&
$N=5$ & \multicolumn{2}{c|}{ \begin{tabular}{c}
  $N=4$ \\
\phantom{$N$}$\heartsuit$ \\
\end{tabular} }  &\begin{tabular}{c}$N=3$ \\
\phantom{$N$}$\heartsuit$ \\
\end{tabular}  &\begin{tabular}{c}  $N=2$ \\
$\heartsuit,\clubsuit$ \\
\end{tabular}  & \begin{tabular}{c}  $N=1$ \\
$\heartsuit,\clubsuit$ \\
\end{tabular} \\
\hline \multicolumn{2}{|c|}{ }   & \multicolumn{7}{c|}{SUGRA}  &
 \multicolumn{2}{c|}{SUGRA/SUSY} & SUGRA & \multicolumn{2}{c|}{SUGRA/SUSY}  \\
\hline
\end{tabular}

\end{center}
\end{table}
\endlandscape
\section{The map: dimensions and supersymmetries} \label{ss:mapsusy}

Table~\ref{tbl:mapsusy} gives an overview on supersymmetric theories in
Minkowski spacetimes and with positive definite kinetic terms. The most
relevant source in this respect is the paper of
Strathdee~\cite{Strathdee:1987jr} that analyses the representations of
supersymmetries.

Supersymmetric field theories of this type in 4 dimensions are restricted
to fields with spins $\leq 2$. This is the restriction to $N\leq 8$ or up
to 32 supersymmetries as an elementary (real) spinor in 4 dimensions has
4 components, see table~\ref{tbl:spinors}. The same table shows that one
can not have more than 11 dimensions if the supersymmetries are
restricted to 32 (at least in spacetimes of Minkowski
signature)~\cite{Nahm:1978tg} . We are considering here the
supersymmetries that square to general coordinate transformations. Thus,
not e.g.\ the special supersymmetries in the superconformal algebra,
which have a different role. The 11-dimensional
theory~\cite{Cremmer:1978km} is the basis of `M-theory', and is therefore
indicated as M in the table.

Going down vertically in the table is obtained by Kaluza-Klein reduction.
That means that one splits all the fields in representations of the lower
dimensional Lorentz group. E.g. the spinors of the 11-dimensional
M-theory split in one right-handed and one left-handed spinor. This is
the theory of the massless sector of IIA string theory, and that is why
we have indicated it as
IIA~\cite{Campbell:1984zc,Huq:1985im,Giani:1984wc}. The massless sector
of IIB theory involves doublets of spinors of the same chirality (thus
also with 32 real supersymmetries). That
theory~\cite{Schwarz:1983wa,Schwarz:1983qr,Howe:1984sr} is not the
dimensional reduction of an 11-dimensional theory, as indicated by its
place in the table.

Elementary supersymmetric theories in 10 dimensions involve only 16
supersymmetries. They appear in superstring theories with open and closed
strings. Apart from the supergravity multiplet which involves the
graviton, one can also have a vector multiplet. The existence of the
simplest matter multiplets (representations of supersymmetry that do not
involve the graviton) is indicated in the table. Vector multiplets
involve a vector, an elementary spinor and possibly scalars. Tensor
multiplets involve antisymmetric tensors $A_{\mu \nu }$. As explained in
section~\ref{ss:bosIngred}, for dimensions 5 and 4 these are dual to
vectors and scalars. Therefore, tensor multiplets are not explicitly
indicated for these lower dimensions. Also, various representations of
the same physical (on-shell) theory are not indicated. The non-Abelian
aspects are not indicated either. E.g.\ the tensor multiplets in 5
dimensions are only dual to vector multiplets when the gauge group is
Abelian, but we do not indicate it separately here.

One can have theories with these matter multiplets also for ungauged
supersymmetry, i.e.\ `rigid supersymmetry'. Rigid supersymmetry is only
possible with up to 16 supersymmetries. Again, this can be understood in
4 dimensions, because for $N>4$ one needs fields of spin-$\ft32$. The
latter are in field theory only possible when they are gauge fields of a
local supersymmetry (gravitinos), which then need gravitons for the gauge
fields of the translations that appear in the commutator of two
supersymmetries.

In the dimensions lower than 10 we indicate the theories by a number $N$
that indicates the symmetry group rotating different supersymmetries. The
structure will be shown explicitly in the next section. For 6 dimensions,
as in 10 dimensions, there are spinors of different chirality, and one
has to distinguish the number of left and right-handed spinors. The
simplest case is with 8 supersymmetries. They have to be all of the same
chirality. The theory is then called $(1,0)$ and would be $N=2$ in the
terminology used in 5 dimensions. With 16 supersymmetries, one can have
$(1,1)$ or $(2,0)$. These are the analogues of IIA and IIB, respectively,
in 10 dimensions. For more supersymmetries, there is a subtlety. The
$(2,1)$ and $(2,2)$ theories can be constructed using a metric tensor
$g_{\mu \nu }(x)$. For the $(4,0)$, $(3,1)$~\cite{Townsend:1984xt} or
$(3,0)$ theories, this field is not present~\cite{Hull:2000zn}, but is
replaced by a more complicated representation of the Lorentz group. Thus,
these theories are different in the sense that they are not based on a
dynamical metric tensor.

If one constructs in 4 dimensions a field theory with $N=7$, then it
automatically has an eighth local supersymmetry. That is why it is not
mentioned in the table. Similarly if one constructs a rigid
supersymmetric theory with $N=3$, it automatically has a fourth
supersymmetry. However, in this case, there is the possibility of having
only three of the four local. Thus $N=3$ is only meaningful in
supergravity. This explains the lowest line of the table.

Finally, let me remark that the vectors of the supergravity theories can
be gauge vectors for a gauge symmetry that rotates the supersymmetries.
The last years, various new results have been obtained in this
direction~\cite{Gheerardyn:2001jj,Hull:2002wg,Bergshoeff:2002mb,%
Andrianopoli:2002mf,Hull:2002cv,Bergshoeff:2002nv,Alonso-Alberca:2002tb,%
deWit:2002vt}. A complete catalogue of theories is not yet known.
However, we believe that all supersymmetric field theories (with a finite
number of fields and field equations that are at most quadratic in
derivatives) belong to one of the entries in table~\ref{tbl:mapsusy}.

\section{Step 1: Supersymmetry and gauge algebra} \label{ss:salgebra}
After this overview of possibilities, we will now give elementary aspects
of the construction of supersymmetric theories. The first basic concept
is the symmetry group.

First, we clarify the relation between transformations and generators. We
have already written transformations of fields, e.g.\
in~(\ref{susyscalar}). This change of a field, is proportional to a
parameter $\epsilon $, and we can write\footnote{We sometimes use spinor
indices $\alpha ,\ldots $ in this section. For details on the notation,
see~\cite{VanProeyen:1999ni}.}
\begin{equation}
  \delta (\OliveGreen{\epsilon })=\OliveGreen{\epsilon ^\alpha}
\Blue{Q_\alpha}\,,
 \label{deltaEpsQ}
\end{equation}
i.e.\ the product of the parameter with an operation called the generator
of the (super)symmetry. This operation is for supersymmetry also a
fermionic object, such that the elementary change of a field is of the
same type as the field itself. When one calculates a commutator of two
transformations, one obtains an anticommutator of the generators:
\begin{eqnarray}
\delta(\OliveGreen{\epsilon _1})\,\delta (\OliveGreen{\epsilon _2})&=&
 \OliveGreen{\epsilon_1 ^\alpha} \Blue{Q_\alpha}\OliveGreen{\epsilon_2 ^\beta}
 \Blue{Q_\beta}=\OliveGreen{\epsilon_2 ^\beta}\OliveGreen{\epsilon_1 ^\alpha} \Blue{Q_\alpha}
 \Blue{Q_\beta}\,, \nonumber\\
\left[ \delta(\OliveGreen{\epsilon _1}),\,\delta (\OliveGreen{\epsilon
_2})\right]&=& \OliveGreen{\epsilon_2 ^\beta}\OliveGreen{\epsilon_1
^\alpha} \Blue{Q_\alpha} \Blue{Q_\beta}-\OliveGreen{\epsilon_1 ^\alpha
}\OliveGreen{\epsilon_2 ^\beta } \Blue{Q_\beta } \Blue{Q_\alpha }
= \OliveGreen{\epsilon_2 ^\beta}\OliveGreen{\epsilon_1
^\alpha}\left(\Blue{Q_\alpha} \Blue{Q_\beta}+\Blue{Q_\beta }
\Blue{Q_\alpha }\right).
\end{eqnarray}
\subsection{Minimal and extended superalgebras}
The minimal supersymmetry algebra is the one that we saw
in~(\ref{susyalgscalar}):
\begin{equation}
  \left\{ \Blue{Q_\alpha },\Blue{Q_\beta }\right\} =
  \gamma ^\mu _{\alpha \beta }\OliveGreen{P_\mu
  }\,.
 \label{minSUSYalg}
\end{equation}
The supersymmetries commute with translations and are a spinor of Lorentz
transformations:
\begin{equation}
 \left[  \OliveGreen{P_\mu },\Blue{Q}\right] =0\,,\qquad
 \left[\OliveGreen{M_{\mu \nu }},\Blue{Q}\right] =-\ft14\gamma _{\mu \nu
 }\Blue{Q}\,.
 \label{QPQM}
\end{equation}
The extensions indicated in the previous section by $N>1$ mean that there
are different supersymmetries $Q^i$ with $i=1,\ldots ,N$. The
possibilities for extension of~(\ref{minSUSYalg}) depend on the reality
properties of the spinors, which were discussed in
section~\ref{ss:Clifford}. In 4 dimensions, with Majorana spinors, and in
5 dimensions with symplectic spinors one can have
\begin{eqnarray}
D=4&:&  \left\{ \Blue{Q_\alpha^i},\Blue{Q_{\beta j}}\right\} =\gamma ^\mu
_{\alpha \beta }\delta ^i_j\OliveGreen{P_\mu
  }\nonumber\\
D=5&:&\left\{ \Blue{Q_\alpha^i},\Blue{Q_\beta ^j}\right\} =\gamma ^\mu
_{\alpha \beta }\Omega ^{ij}\OliveGreen{P_\mu }\,,
 \label{extSUSY45}
\end{eqnarray}
where $\Omega ^{ij}$ is an antisymmetric (symplectic) metric. The
symmetries $\OliveGreen{U}^i{}_j$ that rotate the supercharges by $\left[
\OliveGreen{U}^i{}_j,\Blue{Q}^k\right] =\delta ^k_j\Blue{Q}^i$, are
called \emph{R-symmetries}. The $R$-symmetry group is restricted by the
properties of the spinors. This gives
\begin{eqnarray}
 &&D=10\ :\ \SO(N_L)\times \SO(N_R)\,,\qquad D=9\ :\ \SO(N)\,,\nonumber\\
&& D=8\mbox{ and }D=4\ :\ \U(N)\label{Rsymmetry} \\
  & & D=7\mbox{ and }D=5\ :\ \USp(N)\,,\qquad D=6\ :\ \USp(N_L)\times
  \USp(N_R)\,.\nonumber
\end{eqnarray}
\subsection{4 generalizations}
Apart from these minimal possibilities, one can consider four kinds of
generalizations. The first one is the possibility of central charges, as
found in the classical work of
Haag--{\L}opusza\'nski--Sohnius~\cite{Haag:1975qh}. The simplest example
is in $N=2$, where~(\ref{extSUSY45}) can be extended to
\begin{equation}
  \left\{ \Blue{Q}^i_\alpha ,\Blue{Q}^j_\beta \right\} =
  \gamma ^\mu
_{\alpha \beta }\delta ^i_j\OliveGreen{P_\mu
  }+\varepsilon ^{ij}\left[ {\cal C}_{\alpha \beta }\Red{Z_1}+
  (\gamma _5)_{\alpha \beta }\Red{Z_2}\right].
 \label{D4N2Z}
\end{equation}
The generators $Z_1$ and $Z_2$ commute with everything else and are thus
really `central'. They play an important role when looking for
supersymmetric solutions of the theory. But the name `central charges'
has been generalized to include other generators that can appear in the
anticommutator of supersymmetries. E.g.\ in $D=11$ the properties of the
spinors allow us to extend the anticommutator as~\cite{vanHolten:1982mx}
\begin{equation}
  \left\{\Blue{Q}_\alpha ,\Blue{Q}_\beta  \right\}=\gamma^\mu _{\alpha \beta } \OliveGreen{P}_\mu
+\gamma ^{\mu \nu } _{\alpha \beta }\Red{Z}_{\mu \nu }+\gamma
^{\mu_1\cdots \mu_5 } _{\alpha \beta }\Red{Z}_{\mu_1\cdots \mu_5 }\,.
 \label{QQD11}
\end{equation}
The allowed structures on the right-hand side are determined by the last
entry in table~\ref{tbl:spinors} (remember that this is modulo 4, which
thus allows the 5-index object). The `central charges' $\Red{Z}$ are no
longer Lorentz scalars, and thus do not commute with the Lorentz
generators. They are therefore not `central' in the group-theoretical
meaning of the word, but play in the physical context the same role as
the ones in~(\ref{D4N2Z}), and therefore got the same name.

A second generalization is the extension of the Poincar{\'e} group to the
(anti) de Sitter group. The spacetime with a cosmological constant is a
curved space, which means that translations do not commute, but satisfy
an algebra
\begin{equation}
  \left[ \OliveGreen{P_\mu },\OliveGreen{P_\nu }\right] =\mp \frac{1}{2R^2}\OliveGreen{M_{\mu \nu
  }}\,,
 \label{AdSalgebra}
\end{equation}
where $R$ is related to the inverse of the cosmological constant. The
sign in~(\ref{AdSalgebra}) determines the sign of the cosmological
constant, and, correspondingly, whether the algebra of translations and
Lorentz rotations is $\SO(D,1)$ or $\SO(D-1,2)$. The first one is the de
Sitter algebra, while the second one is the anti-de Sitter algebra.
Extending the first one to a superalgebra needs a non-compact R-symmetry
group, which in turn needs negative kinetic terms of some of the fields,
an undesirable feature. But supersymmetric extensions of anti-de Sitter
algebras are well-known, see the classical work of
Nahm~\cite{Nahm:1978tg} or a recent investigation
in~\cite{D'Auria:2000ec}.

The third generalization is to (super)conformal algebras. The conformal
group is the group consisting of translations $\Blue{P_\mu}$, Lorentz
rotations $\Blue{M_{\mu \nu }}$, dilatations $\Blue{D}$ and special
conformal transformations $\Blue{K_\mu}$, which combine to $\SO(D,2)$. If
one extends it with supersymmetries $\Red{Q^i}$, the algebra requires for
consistency new `special supersymmetries' $\Red{S^i}$ (in the commutator
$[Q,K]$), and the R-symmetry group mentioned above appears in the
anticommutator of the $\Red{Q}$ and $\Red{S}$ supersymmetries.

The fourth generalization is to include extra `Yang--Mills' (YM) gauge
symmetries. Indeed, the spin-1 fields that appear either in the
supergravity multiplet or in the extra vector multiplets, may gauge YM
symmetries according to the principles expressed by~(\ref{covder}). When
the replacement of derivatives by covariant derivatives is performed
everywhere, it is clear that one will not obtain the commutator as
in~(\ref{susyalgscalar}), but rather
\begin{equation}
  \left[ \delta_Q (\OliveGreen{\epsilon _1}),\delta_Q (\OliveGreen{\epsilon _2})\right] \Blue{\phi}
  =\OliveGreen{\bar \epsilon _1}\gamma
^\mu \OliveGreen{\epsilon _2}\,D_\mu \Blue{\phi}
 =\OliveGreen{\bar \epsilon _1}\gamma
^\mu \OliveGreen{\epsilon _2}\,\partial _\mu \Blue{\phi}-\OliveGreen{\bar
\epsilon _1}\gamma ^\mu \OliveGreen{\epsilon _2}\, \Blue{A_\mu ^I} T_I
\Blue{\phi}\,.
 \label{SUSYalgGauge}
\end{equation}
The last term shows in fact a gauge transformation with parameter
$\OliveGreen{\bar \epsilon _1}\gamma ^\mu \OliveGreen{\epsilon _2}\,
\Blue{A_\mu ^I}$. Therefore, $\gamma ^\mu_{\alpha \beta } \Blue{A_\mu
^I}$ is a field-dependent \emph{structure function}, rather than a
structure constant. This type of algebra structure appears often in
supersymmetry, and is called a \emph{soft algebra}. The (adapted) Jacobi
identities then imply that more modifications to the algebra are
necessary. E.g.\ in $N=2$, where the vector multiplets contain scalars
$\sigma ^I$, the algebra has an extra term
\begin{equation}
  \left[ \delta_Q (\OliveGreen{\epsilon _1}),\delta_Q (\OliveGreen{\epsilon _2})\right] \Blue{\phi} =\ldots +
\OliveGreen{\bar \epsilon _1}\OliveGreen{\epsilon _2}\Blue{\sigma ^I}
T_I\Blue{\phi}\,.
 \label{QQextraT}
\end{equation}
When the fields $\sigma ^I$ have a non-zero value for a solution of the
theory, then the algebra that preserves this solution has the form
of~(\ref{D4N2Z}), i.e.\ with central charges.

This illustrates how the first generalization that we discussed above
appears in solutions of the supergravity theories. Also the second
generalization, super-anti-de Sitter algebras, occurs in solutions. The
third and fourth generalization, on the other hand, are important in
constructing supergravity theories.
\subsection{Constructions}\label{ss:construction}
In constructing rigid supersymmetric theories, we start from a rigid
Poincar{\'e} supersymmetry, i.e.\ the Poincar{\'e} symmetries, supersymmetry and
its R symmetries. Then a YM gauge algebra can be added\footnote{In
principle also gauge symmetries of antisymmetric tensors can be included,
which may also have an action on the other fields. We neglect these here
for simplicity, but the principles are the same as for the YM
symmetries.}, gauged by vectors, which are parts of a `vector multiplet'.
The algebra becomes then soft and the rigid supersymmetries mix with the
local YM symmetries as shown in~(\ref{SUSYalgGauge}). In some cases the
action can be invariant under (rigid) superconformal symmetries. Central
charges are not introduced by hand, but may appear due to the mixing of
supersymmetry with YM symmetries and non-zero vacuum expectation values
of some fields, see~(\ref{QQextraT}).

To construct supergravity theories, there is first a straightforward way
(super-Poincar{\'e} way). One gauges gravity and supersymmetry. The
invariance requirements determine all the terms as well in the Lagrangian
as in the transformation laws. The alternative way, called superconformal
tensor calculus, is particularly useful to construct theories with matter
couplings, i.e.\ where `matter multiplets' are coupled to supergravity.
The super-Poincar{\'e} construction leads to a lot of extra terms, which can
be better understood in the context of the superconformal tensor
calculus. The latter starts by gauging the full superconformal algebra.
This is a generalization of the method used to gauge the Poincar{\'e} group
at the end of section~\ref{ss:bosIngred}: gauge fields are associated to
all the symmetries [as in~(\ref{gravitygaugef})], but some of these are
dependent fields using constraints on curvatures [as
in~(\ref{constrRP})]. Then, an action is constructed that is invariant
under the superconformal symmetries, but the symmetries that are
superfluous are gauge-fixed. This is similar to the construction of the
Lagrangian for a massive vector, where in~(\ref{L10}) we constructed an
action invariant under a gauge symmetry, which was then fixed to obtain
the massive vector action in~(\ref{Lmassive1}). For the superconformal
symmetry, this means that dilatations, special conformal transformations,
special supersymmetry and the R-symmetry should be gauge-fixed to obtain
an action that is invariant under the super-Poincar{\'e} group. We thus need
fields that will not be physical [as the scalar $\phi $ in~(\ref{L10})]
but are \emph{compensating} for the symmetries in the superconformal
algebra that do not belong to the super-Poincar{\'e} algebra. They are part
of \emph{compensating multiplets} as they have to be in supersymmetric
representations.

Also the fourth generalization has to be considered in the construction
of general actions: vectors in the supergravity multiplet and vector
multiplets can gauge an extra YM gauge group. In the super-Poincar{\'e}
construction, one considers separately the R-symmetries and other YM
symmetries. One has to define the action of R-symmetry on all the fields
in the theory. In the superconformal tensor calculus, only vectors of
vector multiplets can gauge an extra gauge symmetry $G$, commuting with
supersymmetry. But one of these vector multiplets may be the compensating
multiplet mentioned above. That means that some of its partners are
fields that disappear by gauge conditions. The vectors are the extra
vectors that appear in the `gravity multiplet' from the super-Poincar{\'e}
theory. Also, the R-symmetry is already gauged in the superconformal
context as it is part of the superconformal group, but is afterwards
gauge-fixed. However, the gauge fixing condition may be not invariant
under some of the extra YM gauge symmetries, mixing R-symmetry and $G$.
Schematically we have
\begin{eqnarray}
 \mbox{superconformal including R} & \times  & G \nonumber\\
   & \Downarrow & \mbox{gauge fixing}\nonumber\\
   \mbox{super-Poincar{\'e}} &\mbox{with}& G'\,,
 \label{mixingSPSC}
\end{eqnarray}
where $G'$ is $G$ with a mixing of superconformal R-symmetries. Therefore
(part of) $G'$ does not commute with the supersymmetries and acts as a
gauged R-symmetry.

\section{Step 2: Multiplets and their transformations}
\label{ss:multiplets}

We did already encounter multiplets in the previous sections, especially
in the overview section~\ref{ss:mapsusy}. We now give some more details,
distinguishing on- and off-shell multiplets, and then giving the
properties of the unique multiplets when there are 32 supersymmetries,
the vector multiplets and the multiplets with spins $(\ft12,0)$ for lower
number of supersymmetries.

We first explain the concept of \emph{trivial symmetries}. Consider the
simple action for 2 scalar fields and a gauge invariance:
\begin{equation}
  S=\int
  \rmd x\left[\ft12\Blue{\phi^1}\bbox\Blue{\phi^1}+\ft12\Blue{\phi^2}\bbox\Blue{\phi^2}\right],\qquad
  \delta_{\rm triv} \Blue{\phi^1 }=\OliveGreen{\epsilon}\bbox\Blue{\phi^2}\,, \qquad
 \delta_{\rm triv} \Blue{\phi^2 }=-\OliveGreen{\epsilon}\bbox\Blue{\phi^1}\,.
 \label{symm2scal}
\end{equation}
This can be generalized for any action, when we define transformations
\begin{equation}
  \delta_{\rm triv} \Blue{\phi^i }=\OliveGreen{\epsilon }\,\Red{\eta ^{ij}}\frac{\delta S}{\delta \Blue{\phi
  ^j}}\,,
 \label{DelTriv}
\end{equation}
for any antisymmetric tensor $\Red{\eta ^{ij}}$. Indeed, the variation of
the action is then
\begin{equation}
  \delta_{\rm triv} S=\frac{\delta S}{\delta \Blue{\phi ^i}}\,\OliveGreen{\epsilon }\,\Red{\eta ^{ij}}\frac{\delta S}{\delta \Blue{\phi ^j}}
=0\quad \mbox{if}\quad \Red{\eta ^{ij}}=\Red{-\eta ^{ji}}\,.
 \label{delStriv}
\end{equation}

The relevance of these trivial symmetries is already evident in the
simplest multiplet in $D=4$, $N=1$ supersymmetry: the chiral multiplet.
The multiplet contains a complex scalar $\Blue{z}$ and a fermion
$\Blue{\chi }$, with (rigid) transformations
\begin{equation}
  \delta(\OliveGreen{\epsilon }) \Blue{z}= \OliveGreen{\bar \epsilon}P_L
  \Blue{\chi}\,, \qquad
 \delta(\OliveGreen{\epsilon })P_L\Blue{\chi }  =  P_L\slashed \partial \Blue{z}\OliveGreen{\epsilon}
 \label{transfosChiralM}
\end{equation}
$P_L$ is the projection defined in~(\ref{PLPR}), where in 4 dimensions
$\gamma _*$ is usually indicated as $\gamma _5$. It is sufficient to give
the transformation of $P_L\chi $, as complex conjugation replaces $P_L$
by $P_R$. The action
\begin{equation}
  S=\int \rmd^4 x \,{\cal L} \qquad \mbox{with} \qquad {\cal L}=\Blue{z^*}\bbox \Blue{z}-\ft12\Blue{\bar \chi }\slashed\partial \Blue{\chi
  }\,,
 \label{Schiral}
\end{equation}
is invariant under these transformations. When we calculate the algebra
of the supersymmetries, we find
\begin{equation}
   \left[ \delta(\OliveGreen{\epsilon_1}),\delta(\OliveGreen{\epsilon_2 })\right] \Blue{z} =
 \OliveGreen{\bar \epsilon_2}\gamma ^\mu \OliveGreen{\epsilon _1}\partial _\mu \Blue{z}
 \,,\qquad
   \left[ \delta(\OliveGreen{\epsilon_1}),\delta(\OliveGreen{\epsilon_2 })\right] \Blue{\chi }  =
  \OliveGreen{\bar \epsilon_2}\gamma ^\mu \OliveGreen{\epsilon _1}\left(\partial _\mu\Blue{\chi }
  -\ft12\gamma _\mu \slashed\partial \Blue{\chi }\right).
 \label{algChiral}
\end{equation}
The first terms on the r.h.s. give the general coordinate transformations
and thus represent the minimal supersymmetry algebra. The final term is
proportional to $\slashed\partial \Blue{\chi }$, which is the field
equation of the fermion. This term is of the form of~(\ref{DelTriv}). We
thus find that the supersymmetry algebra~(\ref{minSUSYalg}) is satisfied
on-shell, i.e.\ when the field equations are imposed.

There is a possible improvement here: we can include \emph{auxiliary
fields} $\Red{h}$ (complex). This means that we modify the transformation
rules and Lagrangian to
\begin{equation}
   \delta(\OliveGreen{\epsilon })P_L\Blue{\chi }  =  P_L\slashed \partial\Blue{z}
\OliveGreen{\epsilon} +P_L\Red{h}\OliveGreen{\epsilon } \,, \quad
  \delta(\OliveGreen{\epsilon })\Red{h}  =  \OliveGreen{\bar \epsilon }
\slashed \partial P_L\Blue{\chi}\,, \qquad {\cal L}=\Blue{z^*}\bbox
\Blue{z}-\ft12\Blue{\bar \chi }\slashed\partial \Blue{\chi
}+\Red{h^*}\Red{h}\,.
 \label{withAuxF}
\end{equation}
The auxiliary field has no physical content as its field equation is
$\Red{h}=0$. However, with this modification the
algebra~(\ref{minSUSYalg}) is realized `off-shell', i.e.\ without the
need of equations of motion. One additional advantage is that one can
still envisage other Lagrangians invariant under the same transformation
laws. Indeed, one can e.g.\ add to the Lagrangian a term
\begin{equation}
  {\cal L}_m=m\left( \Blue{z}\Red{h}+\Blue{z^*}\Red{h^*}-\Blue{\bar \chi \chi
  }\right).
 \label{Lchiralm}
\end{equation}
The Lagrangian is still invariant. The field equation of the auxiliary
field is now $\Red{h^*}=-m\Blue{z}$, which lead to massive scalar and
spinor fields. Of course, in the formulation without auxiliary fields,
one has to modify the transformation laws to allow such an extension, as
the last term in~(\ref{algChiral}) would not be a field equation any
more. Therefore, when such a formalism with auxiliary fields is possible,
it is certainly easier to handle, also for considering the quantum
theory. The formalism with auxiliary fields can also be obtained from the
concept of superspace and superfields. However, unfortunately, it is not
always possible to obtain such auxiliary fields.

In general, trivial symmetries can be considered as part of the full set
of transformations of the theory, and as such it makes sense to write the
algebra
\begin{equation}
  \left[ \delta (\OliveGreen{\epsilon _1}),\delta (\OliveGreen{\epsilon _2})\right] \Blue{\phi^i}
=\mbox{minimal susy algebra}+\Red{\eta ^{ij}}(\OliveGreen{\epsilon
_1,\epsilon _2})\,\frac{\delta S}{\delta \Blue{\phi ^j}}
 \label{algWithTriv}
\end{equation}
When we can write the algebra without including the trivial symmetries,
the algebra is called a \emph{closed supersymmetry algebra}. If the
trivial symmetries enter the algebra, then it is called an \emph{open
supersymmetry algebra}. In that case the algebra closes when the field
equations are satisfied or when the (infinite set of) trivial symmetries
are included. Thus in this case, the dynamics can be obtained already
from the algebra of supersymmetry transformations, before constructing
the action.

The square of the supersymmetry operation is in the minimal
algebra~(\ref{minSUSYalg}) a general coordinate transformation. This is
an invertible operation. As the supersymmetry operation transforms boson
in fermion states, and vice-versa, one can conclude that the number of
boson and fermion states should be the same. This is thus true when the
algebra of supersymmetries gives just $\OliveGreen{P}$. E.g.\ with the
transformations~(\ref{transfosChiralM}) and algebra~(\ref{algChiral}), we
can apply this only for on-shell states, such that $\slashed\partial
\Blue{\chi }=0$. Then we count 2 bosonic states for the complex
$\Blue{z}$, and the 4 fermionic components of $\Blue{\chi }$ are reduced
to 2 by the field equation. So we have a $2+2$ on-shell multiplet. When
the auxiliary field $\Red{h}$ is included, the algebra is also satisfied
off-shell. Thus we have in this case the 4 components of $\Blue{\chi }$
and $\Blue{z}$ and $\Red{h}$ give together also 4 components. In this
case, we say that we have a $4+4$ off-shell multiplet. These two ways of
counting are called \emph{on-shell counting} and \emph{off-shell
counting}. Let me finally remark that to have the minimal algebra we have
to eliminate the extra gauge terms that we illustrated
in~(\ref{SUSYalgGauge}). Thus, we always have to subtract gauge degrees
of freedom in all countings. E.g., a gauge vector in 4 dimensions would
count off-shell for 3 degrees of freedom, and on-shell for 2 degrees of
freedom [representation of $\SO(D-2)$, see section~\ref{ss:bosIngred}].

We now give some general facts about the most important multiplets in 4
dimensions. The  \emph{pure supergravity} multiplet is the set of fields
that represents the spacetime susy algebra and has gauge fields for the
supersymmetries. The number of fields is given in
table~\ref{tbl:pureSGD4}.
\begin{table}[htbp]
  \caption{\it Pure supergravity multiplets in 4 dimensions according to spin $s$}
  \label{tbl:pureSGD4}
\begin{center}
  $\begin{array}{|c|ccccccc|}\hline
    s & N=1 & N=2 & N=3 & N=4 & N=5 & N=6 & N=8 \\ \hline
    2 & 1 & 1 & 1 & 1 & 1 & 1 & 1 \\
    \ft32 & 1 & 2 & 3 & 4 & 5 & 6 & 8 \\
    1 &   & 1 & 3 & 6 & 10 & 16 & 28 \\
    \ft12 &   &   & 1 & 4 & 11 & 26 & 56 \\
    0 &   &   &   & 2 & 10 & 30 & 70 \\ \hline
  \end{array}$
\end{center}
\end{table}
These are on-shell multiplets. The fields of spin $s>0$ have 2 degrees of
freedom (helicity $+s$ and $-s$). If $N\leq 4$ one can add to the theory
\emph{matter multiplets} with fields $\leq 1$. Those containing a spin-1
field are called vector multiplets, and the multiplets for $N\leq 2$ with
spin $\leq \ft12$ are called hypermultiplets for $N=2$ or the already
illustrated chiral multiplet for $N=1$, see table~\ref{tbl:mattermD4}.
\begin{table}[htbp]
  \caption{\it Matter multiplets in 4 dimensions}\label{tbl:mattermD4}
\begin{center}
  $\begin{array}{|c|ccc|}\hline
    s & N=1 & N=2 & N=3,4 \\ \hline
    1  & 1 & 1 & 1 \\
    \ft12 & 1 & 2 & 4 \\
    0 &   & 2 & 6 \\ \hline
  \end{array}     \qquad\qquad
\begin{array}{|c|cc|}\hline
    s & N=1 & N=2 \\ \hline
    \ft12 & 1 & 2 \\
    0 & 2 & 4 \\ \hline
  \end{array}$
\end{center}
\end{table}
An arbitrary number of these matter multiplets can be used for rigid
supersymmetry or can be added to the gravity multiplet in local
supersymmetry (supergravity). The vectors in the vector multiplets and
those in the gravity multiplets can gauge an extra (possibly non-Abelian)
gauge group. In rigid supersymmetry one can only have compact gauge
groups if one requires positive kinetic energies, but in supergravity
some non-compact gauge groups are possible without spoiling the
positivity of the kinetic energies. However, the list of possible
non-compact groups is restricted for any $N$. A number of hypermultiplets
($N=2$) or chiral multiplets ($N=1$) may then form a representation of
these gauge groups.

When there are 32 supersymmetries, the multiplet is unique and this
multiplet is only known on-shell, i.e.\ it is not known (and there are
no-go theorems) how to add auxiliary fields to obtain off-shell closure.
The basic multiplet in 11 dimensions is written in terms of just 3
fields: a graviton $\Maroon{g_{\mu \nu }}$ (44 components, as traceless
symmetric tensor of $\SO(9)$), an antisymmetric $\Maroon{A_{\mu \nu \rho
}}$ (84 components) and a vector-spinor $\Red{\psi _\mu}$ build the
$128+128$ multiplet. Reducing this e.g.\ to $D=4$ fields gives the $N=8$
multiplet in table~\ref{tbl:pureSGD4}.

With 16 supersymmetries, rigid supersymmetry is possible. One can have
vector multiplets, or tensor multiplets for $(2,0)$ supersymmetry in 6
dimensions. In $D=5$, the tensor multiplets are dual to vector multiplets
at the level of zero gauge coupling constant. Gauging breaks this
duality. Supergravity theories with 16 supersymmetries may contain a
number of these multiplets. The model is fixed once one gives the number
of matter multiplets and the gauging that is performed by the vectors.

Theories with 8 or 4 supersymmetries are not fixed by the discrete
choices of number of multiplets and gauging. In these cases the model
depends on some functions that can vary by infinitesimal variations. It
is in these models that auxiliary fields are most useful. E.g.\ for the
chiral multiplets that we mentioned before, a holomorphic function
$\OliveGreen{W}(\Blue{z})$ can be introduced, which may take arbitrary
values, determining a potential. Indeed, the addition~(\ref{Lchiralm})
can be generalized to
\begin{equation}
  {\cal L}_W= \frac{\partial \OliveGreen{W}(\Blue{z})}{\partial \Blue{z}}\,\Red{h}
-\frac{\partial ^2\OliveGreen{W}(\Blue{z})}{\partial
\Blue{z}^2}\Blue{\bar \chi} P_L \Blue{\chi} +\hc \,.
 \label{LchiralW}
\end{equation}
Similarly, the kinetic terms can be generalized depending on an arbitrary
function ${\cal G}(z,z^*)$, which will play the role of a K{\"a}hler
potential for the geometry determined by the scalars (see
section~\ref{ss:scalGeom}).

\section{Step 3: Actions} \label{ss:actions}

The next step is the determination of the action. There are some general
properties of actions that we will show in this section. The kinetic
terms of the scalars determine a geometry. There is also a potential for
these scalars with properties that are determined by the supersymmetry.
The kinetic terms of the vectors introduce the duality symmetries, which
by supersymmetry imply symmetries on the full theory, leading often to
hidden symmetries of the scalar geometry.

The full action of a supergravity theory is very complicated. It contains
4-fermion couplings, couplings between fermions and vectors (as dipole
moments), \ldots . We show here some general structure of the bosonic
terms in 4 dimensions. The theory contains\footnote{We neglect here the
possibility of additional antisymmetric tensors, though it is not proven
whether all theories with antisymmetric tensors have an equivalent
description in terms of scalars.} the graviton, represented by the
vierbein $e_\mu ^a$, a number of vectors $A_\mu ^I$ with field strengths
$\mathcal{F}_{\mu \nu }^I$, a number of scalars $\Red{\varphi ^u}$, $N$
gravitinos $\Blue{\psi^i _\mu}$, and a number of fermions $\Blue{\lambda
^A}$. The pure bosonic terms of such an action are
\begin{eqnarray}
 e^{-1}{\cal L}_{\rm bos}  &=& \ft12R+\ft14 (\Red{\Im {\cal N}_{IJ}}){\cal F}_{\mu\nu}^I {\cal F}^{\mu\nu
J} -\ft 18 (\Red{\Re {\cal N}_{IJ}})e^{-1}
\varepsilon^{\mu\nu\rho\sigma}{\cal
F}_{\mu\nu}^I {\cal F}_{\rho\sigma}^J \nonumber\\
&&   -\ft12 \Red{g_{uv}(\varphi)}\,D_\mu \Red{\varphi^u}\, D^\mu
\Red{\varphi^v}
   -\Maroon{V}\Red{(\varphi)}\,. \label{generalbosL}
\end{eqnarray}
We factorized the determinant of the vierbein to the left hand side. The
first term gives the pure gravity action. Then there are the kinetic
terms for the spin-1 fields. They depend on two tensor functions, which
we combine in a complex (symmetric) tensor $\Red{{\cal N}_{IJ}}$. This
tensor in general is a function of the scalars in the theory. The scalars
have kinetic terms determined by a symmetric tensor
$\Red{g_{uv}(\varphi)}$. They couple `minimally' to the vectors with a
covariant derivative for the gauge symmetries as in~(\ref{covder}).
Finally, there is the potential $\Maroon{V}\Red{(\varphi)}$.

We will now illustrate the duality transformations in $D=4$,
generalizations of the Maxwell dualities\footnote{Aspects of dualities
and gaugings in arbitrary dimensions are treated in~\cite{Fre:2001jd}.}.
They were first discussed
in~\cite{Ferrara:1977iq,deWit:1979sh,Cremmer:1979up,Gaillard:1981rj}.
They apply only for Abelian theories and without the coupling of the
vectors to other fields. Thus, e.g.\ we neglect the appearance of the
vector in covariant derivative of the scalars in~(\ref{generalbosL}). The
vectors then appear in the action only as their field strengths. If we
express the theory in terms of field strengths, we have to complement the
field equations with Bianchi identities $\varepsilon ^{\mu \nu \rho
\sigma }\partial_\mu\mathcal{F}_{\rho \sigma}=0$. But there is a
convenient way to write the Bianchi identities and field equations:
\begin{eqnarray}
&&\partial^\mu \Im {\mathcal F}^{+I }_{\mu\nu} =0\,,\qquad  {\mathcal
F}^{\pm I }_{\mu\nu}\equiv \ft12\left( {\mathcal F}^{ I }_{\mu\nu} \pm
\ft12 \rmi e\varepsilon _{\mu \nu \rho \sigma }\mathcal{F}^{\rho \sigma
\,I}\right) \nonumber\\
&&\partial_\mu \Im G_{+I }^{\mu\nu} =0\,,\qquad  G_{+I }^{\mu\nu}\equiv
2\rmi\frac{\partial{\mathcal L}}
  {\partial {\mathcal F}^{+I }_{\mu\nu}}=
{\mathcal N}_{I J }{\mathcal F}^{+J \,\mu\nu}\,. \label{BianchiFE}
\end{eqnarray}
This shows that, for $m$ vectors ($I=1,\ldots ,m$), this set of equations
is invariant under $G\ell (2m,\mathbb{R})$:
\begin{equation}
  \pmatrix{\tilde \mathcal{F}^+\cr \tilde G_+}= \mathcal{S} \pmatrix{\mathcal{F}^+\cr
  G_+}= \pmatrix{A&B\cr C&D} \pmatrix{\mathcal{F}^+\cr
  G_+}\,.
 \label{Gl2mR}
\end{equation}
These transformations imply a new form of the tensor $\mathcal{N}$:
\begin{eqnarray}
&&  \tilde
G^+=(C+D\mathcal{N})\mathcal{F}^+=(C+D\mathcal{N})(A+B\mathcal{N})^{-1}
  \tilde \mathcal{F}^+\nonumber\\
  && \Rightarrow \ \ \tilde
  \mathcal{N}=(C+D\mathcal{N})(A+B\mathcal{N})^{-1}\,.
 \label{tildeN}
\end{eqnarray}
Consistency with the last of~(\ref{BianchiFE}) implies that $\tilde
\mathcal{N}$ should be symmetric. Writing out these conditions, one
arrives at the conclusion that $\mathcal{S}$ should be in
$\Symp(2m,\mathbb{R})$. Thus the vector field strengths belong to
$2m$-symplectic vectors.

To understand the scalar geometry, we have to distinguish 2 manifolds. On
the one hand there is the spacetime, with coordinates $x^\mu $, $\mu
=0,1,2,3$. On the other hand there is a manifold, which we indicate as
${\cal M}$, with dimension equal to the number of scalars in the model.
The scalars $\Red{\varphi^u}$ give a chart in this manifold. The values
of a scalar at a spacetime point $x$ thus determine a submanifold of
${\cal M}$, parametrized by $\Red{\varphi^u}(x)$. The metric of spacetime
is $g_{\mu \nu }(x)=e_\mu ^a(x) \eta _{ab} e_\nu ^b(x)$. The metric on
${\cal M}$ is $\Red{g_{uv}(\varphi)}$. These have a different status.
While the latter is part of the definition of the theory, $g_{\mu \nu
}(x)$ is, together with $\Red{\varphi^u}(x)$, a dynamical field. On the
other hand, the induced metric on spacetime is $\Red{g_{uv}(\varphi
)}\,(\partial _\mu \Red{\varphi ^u})(\partial _\nu \Red{\varphi ^v})$ at
$\varphi =\varphi (x)$, and its contraction with the (inverse) spacetime
metric and its determinant $\sqrt{g}g^{\mu \nu }$ appears in the action.

The scalar manifold can have isometries, i.e.\ symmetries of the induced
metric $\rmd s^2=\Red{g_{uv}(\varphi )}\,\rmd \Red{\varphi ^u}\,\rmd
\Red{\varphi ^v}$. Usually these symmetries are extended to a symmetry of
the full action (there are counterexamples, but they are rare). This
group is then called the \emph{U-duality group}. The scalars and the
vectors are connected via the tensor $\Red{\mathcal{N}_{IJ}(\varphi )}$.
Therefore the isometries act as duality transformations in the vector
sector, and as such must belong to the $\Symp(2m,\mathbb{R})$ group (in 4
dimensions). This gives a restriction of possible U-duality groups, which
are for $D=4$ restricted to be a subgroup of $\Symp(2m,\mathbb{R})$ for
scalars that belong to multiplets including vectors. Actually, to count
the $m$ vectors, we have to include those of the gravity multiplet. This
is natural in superconformal tensor calculus, where these vectors belong
to vector multiplets of which part of the fields are compensating fields,
see section~\ref{ss:construction}. A subgroup of the isometry group, at
most of dimension $m$, can then be gauged. This means that the vectors
couple to the scalars in the covariant derivative using in~(\ref{covder})
the transformations of the scalars under these isometries. There are more
corrections due to this gauging in the fermionic sector. We will give
more details on the geometries in section~\ref{ss:scalGeom}, but first
finish the overview of the different steps of the analysis of
supersymmetric theories.

\section{Step 4: Solutions and their symmetry} \label{ss:solutions}

We usually look for solutions with vanishing fermions. This is often
motivated by the desire to keep at least some of the Lorentz invariance
unbroken. A non-vanishing fermion is not invariant under any part of the
Lorentz group. The values of the metric, vector fields and scalar fields
then determine the type of solution that we are discussing. These may be
Anti-de Sitter geometries, black holes, branes or pp-waves or Minkowski
spaces, which all preserve some supersymmetry. When we discuss preserved
supersymmetry, this means some rigid supersymmetry. There is often a
confusing terminology that a solution preserves all supersymmetries. What
is meant is that from all the local supersymmetries parametrized by
$\OliveGreen{\epsilon^i_\alpha} (x)$, indicating now as well the spinor
index $\alpha=1,\ldots ,\Delta $ as the index $i=1,\ldots, N$ for the
extension, there are specific functions depending on $\Delta N$ constant
parameters that are invariances of the solution.

To find preserved supersymmetries, we have to consider the
transformations of the form
\begin{equation}
  \delta (\OliveGreen{\epsilon })\,{\rm boson}=\OliveGreen{\epsilon
  }\,\textrm{fermion}\,,\qquad
\delta (\OliveGreen{\epsilon })\,{\rm fermion}=\OliveGreen{\epsilon
  }\,\textrm{boson}\,.
 \label{delbosonfermion}
\end{equation}
For vanishing fermions, we have to consider the condition of vanishing
transformations of the fermions to determine the preserved
supersymmetries. A solution (a bosonic configuration) that allows
non-zero parameters $\OliveGreen{\epsilon }$, is called a BPS solution.
The algebra of supersymmetry implies for most of these solutions a
cancellation between e.g.\ contributions of the energy and of the
electromagnetic (or other) charges. This can be seen already
from~(\ref{SUSYalgGauge}) or~(\ref{QQextraT}). For preserved
supersymmetries, the right-hand side should vanish when applied to a
solution. There are non-zero terms proportional to the energy determined
by $\partial _0\phi $ and proportional to charges determined by $T_I\phi
$. In solutions with non-zero gauge fields (e.g.\ charged black holes)
the last term of~(\ref{SUSYalgGauge}) has to cancel the energy, while for
non-zero scalars, the term in~(\ref{QQextraT}) plays this role. Thus, in
any case these solutions satisfy some bounds on charges that are called
Bogomol'nyi bounds.

This happens e.g.\ for charged black holes in $D=4$, $N=4$
supergravity~\cite{Kallosh:1992ii}. The solutions may have electric (P)
and magnetic (Q) charges. They satisfy $P^2+Q^2\leq M^2$, where $M$ is
the mass of the black hole. This bound is automatic for solutions of the
supersymmetric theories as a consequence of the algebra, and coincides
with the requirement of cosmic censorship (no naked singularities in
spacetime). If there is an equality, then there are solutions for the 16
functions $\OliveGreen{\epsilon^i_\alpha} (x)$ that depend on 4 constant
parameters ($N=1$ in $D=4$). If, moreover, either $P$ or $Q$ is zero,
then there are 8 solutions ($N=2$).

\section{Scalars and geometry} \label{ss:scalGeom}

We finish these lectures by giving an overview of the geometries that
appear in the scalar manifolds, as explained in section~\ref{ss:actions}.
The type of geometries that occur, depend on the number of supercharges.
For all the theories with more than 8 supersymmetries, all the scalar
manifolds are symmetric spaces. These are shown in
table~\ref{tbl:geometriesPlus8}. For the theories with 4 supersymmetries
($N=1$ in 4 dimensions, but one might also consider lower-dimensional
theories), the manifold can be an arbitrary K{\"a}hler geometry, a geometry
with a (closed) complex structure. The symmetric K{\"a}hler spaces are
\begin{eqnarray}
&&\frac{\SU(p,q)}{\SU(p)\times \SU(q)\times \U(1)}  \,,\qquad
\frac{\SO^*(2n)}{\U(n)}\,,\qquad \frac{\Symp(2n)}{\U(n)}\,, \nonumber\\
&& \frac{\SO(n,2)}{\SO(n)\times \SO(2)}\,,\qquad
\frac{\mathrm{E}_6}{\SO(10)\times \U(1)} \,,\qquad
\frac{\mathrm{E}_7}{\mathrm{E}_6\times \U(1)}\,.
 \label{symmKahler}
\end{eqnarray}
Arbitrary K{\"a}hler spaces are defined by a K{\"a}hler potential ${\cal
G}(z,z^*)$, mentioned at the end of section~\ref{ss:multiplets}. For any
K{\"a}hler manifold there is such an $N=1$ theory.

Beautiful structures emerge for theories with 8 supercharges ($N=2$ if in
$D=4$). These theories all belong to a class that was baptized
\emph{special geometries}~\cite{Strominger:1990pd,deWit:1992cr},
including some real~\cite{Gunaydin:1984bi}, some K{\"a}hler
geometries~\cite{deWit:1984pk} and all the quaternionic
geometries~\cite{Bagger:1983tt}. Especially, the scalars that by
supersymmetry are directly related to vectors have a geometrically
distinct structure, special K{\"a}hler geometry~\cite{deWit:1984pk}. This is
a subclass of the K{\"a}hler geometries discussed above, with an extra
symplectic symmetry structure related to the duality transformations of
the vectors shown in section~\ref{ss:actions}. Scalars in hypermultiplets
exhibit quaternionic structures, with many relations with special K{\"a}hler
manifolds~\cite{Cecotti:1989qn,deWit:1993wf}.

Specifically, the manifolds that occur in supergravity actions are
\begin{eqnarray}
  D=6& :& \frac{\mathrm{O}(1,n)}{\mathrm{O}(n)} \times \mbox{quaternionic-K{\"a}hler manifold} \nonumber\\
  D=5&:& \mbox{very special real manifold} \times \mbox{quaternionic-K{\"a}hler manifold} \nonumber\\
  D=4&:& \mbox{special K{\"a}hler manifold} \times \mbox{quaternionic-K{\"a}hler manifold.}
 \label{D654Special}
\end{eqnarray}
A short overview of these manifolds is given in~\cite{VanProeyen:2001wr},
especially in sections~2 and~3, where tables are given of the symmetric
special geometries and homogeneous special geometries. Indeed, the work
on these couplings lead to new mathematical discoveries in the field of
quaternionic geometry~\cite{deWit:1992nm,deWit:1995tf}, especially an
improvement on the classification of homogeneous quaternionic
geometries~\cite{Alekseevsky1975}.

These geometries determine the general couplings of supergravity to
matter multiplets in $D=6$~\cite{Bergshoeff:1986mz,Riccioni:2001bg},
$D=5$~\cite{Ceresole:2000jd} and
$D=4$~\cite{deWit:1985px,Andrianopoli:1997cm}. There exist also versions
of these geometries for rigid supersymmetry, leading to rigid K{\"a}hler
manifolds~\cite{Sierra:1983cc,Gates:1984py} and hyperk{\"a}hler manifolds.

Another new aspect, which has shown recently~\cite{Bergshoeff:2002qk}, is
the possibility of generalization of hyperk{\"a}hler to hypercomplex
manifolds for rigid hypermultiplets and of quaternionic-K{\"a}hler to
quaternionic manifolds for hypermultiplets in supergravity. This
generalization involves theories where no invariant metric can be
defined. Then the field equations do not follow from an action, but are
determined by non-closure functions as in~(\ref{algWithTriv}), but where
the last factor is a dynamical equation $E_j$ that can not be written as
the derivative of some action $E_j\neq \frac{\delta S}{\delta \Blue{\phi
^j}}$.

\section{Final remarks} \label{ss:final}

We know a lot of the general structure of supergravity theories, but
still new aspects of supergravity theories are discovered every day. They
lead to interesting applications in phenomenology and even cosmology
these days. For those who want to study further the aspects of
supergravity theories, we refer to the recent longer review of
B.~de~Wit~\cite{deWit:2002vz}.

 \landscape
\begin{table}[htbp]
  \caption{{\it Scalar geometries in theories with more than 8
  supersymmetries (and dimension $\geq 4$).} The theories are ordered as
  in table~\ref{tbl:mapsusy}. Note that the R-symmetry
  group, mentioned in~(\ref{Rsymmetry}), is always a factor in the
  isotropy group. For more than 16 supersymmetries, there is only a
  unique supergravity (up to gaugings irrelevant to the geometry),
  while for 16 and 12 supersymmetries there is a number $n$ indicating
  the number of vector multiplets that are included.}\label{tbl:geometriesPlus8}
\begin{center}
  $\begin{array}{|@{\hspace{2pt}}c| 
  *{3}{c|} @{\hspace{2pt}}c| 
c|@{\hspace{2pt}}c| c| *{2}{@{\hspace{2pt}}c|} *{1}{c|}
   }
\hline
 D & \multicolumn{4}{c|}{32} & \multicolumn{2}{c|}{24} &20  & \multicolumn{2}{c|}{16} & 12      \\
\hline
 10 & {\mathrm O}(1,1)&\frac{\SU(1,1)}{\OliveGreen{\U(1)}}& \multicolumn{2}{c|}{ }&\multicolumn{2}{c|}{ }&&&&\\
 9  & \multicolumn{2}{c|}{\frac{\Sl(2)}{\OliveGreen{\SO(2)}} \otimes
  {\mathrm O}(1,1)}&\multicolumn{2}{c|}{ }&\multicolumn{2}{c|}{ }
&& \frac{\mathrm{O}(1,n)}{\mathrm{O}(n)} \otimes  {\mathrm O}(1,1)& &    \\[5mm]
8  & \multicolumn{2}{c|}{\frac{\Sl(3)}{\OliveGreen{\SU(2)}}\otimes
\frac{\Sl(2)}{\OliveGreen{\U(1)}} }&\multicolumn{2}{c|}{
}&\multicolumn{2}{c|}{ }  &  & \frac{{\mathrm
O}(2,n)}{\OliveGreen{\U(1)}\times
  {\mathrm O}(n)}\otimes  {\mathrm O}(1,1)&  &      \\[5mm]
7  & \multicolumn{2}{c|}{\frac{\Sl(5)}{\OliveGreen{\USp(4)}} }&
\multicolumn{2}{c|}{ }&\multicolumn{2}{c|}{ } & & \frac{{\mathrm
O}(3,n)}{\OliveGreen{\USp(2)}\times
  {\mathrm O}(n)}\otimes  {\mathrm O}(1,1)& &      \\[5mm]
6  & \multicolumn{2}{c|}{\frac{\mathrm{O}(5,5)}{\OliveGreen{\USp(4)\times
\USp(4)}}}&\frac{\textrm{F}_4}{\OliveGreen{\USp(6)\times
\USp(2)}}&\frac{\textrm{E}_6}{\OliveGreen{\USp(8)}}
&\frac{\SU^*(4)}{\OliveGreen{\USp(4)}}&\frac{\SU^*(6)}{\OliveGreen{\USp(6)}}
& & \frac{\mathrm{O}(4,n)}{\mathrm{O}(n)\times \OliveGreen{SO(4)}}\otimes
{\mathrm O}(1,1)
 & \frac{\mathrm{O}(5,n)}{\mathrm{O}(n)\times \OliveGreen{\USp(4)}} &     \\[5mm]
5  & \multicolumn{4}{c|}{\frac{\mathrm{E}_6}{\OliveGreen{\USp(8)}}} &
\multicolumn{2}{c|}{\frac{\SU^*(6)}{\OliveGreen{\USp(6)}}} &  &
\multicolumn{2}{c|}{\frac{{\mathrm O}(5,n)}{\OliveGreen{\USp(4)}\times
  {\mathrm O}(n)}\otimes  {\mathrm O}(1,1)}   &      \\[4mm]
4  & \multicolumn{4}{c|}{\frac{\mathrm{E}_7}{\OliveGreen{\SU(8)}}} &
\multicolumn{2}{c|}{\frac{SO^*(12)}{\OliveGreen{\U(6)}}} &
\frac{\SU(1,5)}{\OliveGreen{\U(5)}} &
\multicolumn{2}{c|}{\frac{\SU(1,1)}{\OliveGreen{\U(1)}}\times
\frac{\SO(6,n)}
{\OliveGreen{\SU(4)}\times \SO(n)}}  & \frac{\SU(3,n)}{\OliveGreen{\U(3)}\times \SU(n)}    \\
\hline
\end{array}$
\end{center}
\end{table}
\endlandscape

%
%

\providecommand{\href}[2]{#2}\begingroup\raggedright\endgroup

\label{endxyzt}
\end{document}